\documentclass[useAMS,usenatbib]{mn2e}

\title[RRATs and pulsar nulls]{An incomplete model of RRATs and of nulls %%@
mode-changes and subpulses}
\author[P. B. Jones]{P. B. %%@
Jones\thanks{E-mail:p.jones1@physics.ox.ac.uk}\\
Department of Physics, University of Oxford, Denys Wilkinson Building,\\
Keble road, Oxford OX1 3RH, England}
\begin{document}

\date{}

\pagerange{\pageref{firstpage}--\pageref{lastpage}} \pubyear{}

\maketitle

\label{firstpage}

\begin{abstract}
A model for pulsars with polar-cap magnetic flux density ${\bf B}$ %%@
antiparallel with spin ${\bf \Omega}$ is described.  It recognizes the %%@
significance of two elementary processes, proton production in %%@
electromagnetic showers and photoelectric transitions in ions accelerated %%@
through the blackbody radiation field, which must be present at the polar %%@
cap in the ${\bf \Omega}\cdot{\bf B} < 0$ case, but not for pulsars of %%@
the opposite spin direction.  The two populations are likely to be %%@
indistinguishable observationally until curvature radiation pair creation %%@
ceases to be possible.  The model generates, and provides a physically %%@
realistic framework for, the polar-cap potential fluctuations and their %%@
time-scales that can produce mode-changes and nulls.  The RRATs are then %%@
no more than an extreme case of the more commonly observed nulls.  The %%@
model is also able to support the basic features of subpulse drift and to %%@
some extent the null-memory phenomenon that is associated with it.  %%@
Unfortunately, it appears that the most important neutron-star parameter %%@
for quantitative predictive purposes is the whole-surface temperature %%@
$T_{s}$, a quantity which is not readily observable at the neutron-star %%@
ages concerned.
\end{abstract}


\section[]{Introduction}

The discovery within the last decade of further complex pulsar phenomena, %%@
particularly the Rotating Radio Transients (RRATs; McLaughlin et al %%@
2006), has extended the problem of understanding these systems by %%@
introducing new time-scales.  In a number of previous papers (Jones %%@
2010a, 2011, 2012a, 2012b; hereafter Papers I-IV) physical processes at %%@
the polar-cap surface in pulsars with spin ${\bf \Omega}$ and magnetic %%@
magnetic flux density ${\bf B}$ such that ${\bf \Omega}\cdot{\bf B} < 0$ %%@
have been examined to see if they are relevant. The processes that %%@
introduce important time-scales are those associated with the formation %%@
of electromagnetic showers by reverse-electrons from either %%@
electron-positron pair formation or from photoelectric transitions in %%@
accelerated ions. They are of no significance in ${\bf \Omega}\cdot{\bf %%@
B} > 0$ pulsars with Goldreich-Julian charge density $\rho _{GJ} < 0$ and %%@
outward electron acceleration.  The work has assumed that the %%@
space-charge-limited flow (SCLF) boundary condition ${\bf E}\cdot{\bf B} %%@
= 0$ is satisfied on the polar-cap surface at all times.

The aim of these papers has been to determine the composition and energy %%@
distribution of the accelerated particle flux and to look for %%@
instabilities that might be relevant to the phenomena which are observed %%@
in the radio-frequency spectrum. These properties can then be compared %%@
with those needed to produce the observed emission spectra. This has a %%@
large bandwidth, of the order of $10^{9}$ Hz, and efficient conversion of %%@
kinetic energy to radio frequencies.  For example,
the peak power of the brightest pulses in PSR B0656+14 (Weltevrede et al %%@
2006a), is of the order of $7\times 10^{27}$ erg s$^{-1}$, which is %%@
equivalent to approximately $10^{2-3}$ MeV per unit charge accelerated at %%@
the polar caps.

Growth of a collective mode able to transfer energy at this rate to the %%@
radiation field constrains the longitudinal effective mass, $m_{i}\gamma %%@
^{3}_{i}$, of beam particles with mass $m_{i}$ and Lorentz factor $\gamma %%@
_{i}$. For a secondary electron-positron pair plasma, it is well known %%@
that this requires a low energy, $\gamma _{e}\sim 100$, which is also of %%@
the same order as that needed if coherent curvature radiation, in a %%@
dipole-field, were the source of the observed radiation.  Papers III and %%@
IV showed that, under the SCLF boundary condition, the creation of a %%@
reverse flux of electrons by photoelectric transitions in the accelerated %%@
ions limits the acceleration potential, analogously with the effect of %%@
electron-positron pair creation. Thus the outward particle flux has two %%@
principal components: protons formed in the electromagnetic showers and %%@
ions with Lorentz factors $\gamma _{p}\approx 2\gamma _{A,Z}$ which are %%@
relativistic, but not ultra-relativistic as they would be in the absence %%@
of photoelectric transitions.
They can have longitudinal effective masses that allow the rapid growth %%@
of the quasi-longitudinal Langmuir mode considered by Asseo, Pelletier \& %%@
Sol (1990).
As noted by Asseo et al, the quasi-longitudinal mode couples directly %%@
with the radiation field and so introduces, in principle, a second source %%@
of coherent radio-frequency emission which is not present in ${\bf %%@
\Omega}\cdot{\bf B} > 0$ pulsars.

Electromagnetic shower theory (see Landau \& Rumer 1938, Nordheim \& Hebb %%@
1939)
makes it possible to calculate the total photon track length per unit %%@
interval of photon frequency. It is an almost linear function of primary %%@
electron energy in the ultra-relativistic limit.  Then known partial %%@
cross-sections for the formation and decay of the giant dipole resonance %%@
enable $W_{p}$, the number of protons formed per unit incident electron %%@
energy, to be estimated with adequate reliability for the present work.  %%@
Proton formation is concentrated at shower depths of the order of $10$ %%@
radiation lengths and it was shown in Paper I that the time $\tau _{p}$ %%@
for diffusion to the top of the polar-cap atmosphere introduces the %%@
likelihood of instability in the composition of the accelerated particle %%@
flux. Estimates in Paper II, under more appropriate assumptions about the %%@
depth of the atmosphere, were that $\tau _{p}$ would be of the order of %%@
$10^{-1}-10^{0}$ s. With this result in mind, it was evident that the %%@
photoelectric transitions considered in Papers III and IV would introduce %%@
large-scale fluctuations in the the acceleration potential above the %%@
polar cap.  The time-scale $\tau _{p}$ is sufficiently long that the %%@
electric field is simply given by solving Poisson's equation for the %%@
charge densities instantaneously present above the polar cap.  It follows %%@
that the conditions necessary for coherent radio emission can appear or %%@
disappear with time-scales related to $\tau _{p}$.

Before proceeding further, it is appropriate to place this paper in the %%@
context of recent studies of the pulsar magnetosphere based on %%@
computational techniques in numerical plasma kinetics.  It is true that
the inductance per unit length of the tube of open magnetic flux lines %%@
varies little with altitude above the polar cap so that the magnetosphere %%@
as a whole limits current-density time derivatives.   The further %%@
hypothesis that the actual current densities are determined by the whole %%@
magnetosphere rather than by considerations specific to the small region %%@
immediately above the polar cap appeared in a paper by Mestel et al %%@
(1985) and was later extended by Beloborodov (2008).  Solutions for the %%@
magnetic flux density external to a sphere, typically of radius $\sim %%@
0.2R_{LC}$, where $R_{LC}$ is the light-cylinder radius, have been %%@
obtained by Kalapotharakos \& Contopoulos (2009) and by Bai \& Spitkovsky %%@
(2010) under the assumption of a force-free magnetosphere. Current %%@
densities can then be derived from the ${\bf B}$-distribution as %%@
functions of basic parameters; the dipole moment, the rotation period %%@
$P$, and the angle $\psi$ subtended by ${\bf \Omega}$ and ${\bf B}$.  %%@
Those found by Bai \& Spitkovsky deviate considerably from
${\rho _{GJ}}c$, the Goldreich-Julian value, and were followed by %%@
one-dimensional time-dependent numerical studies of the polar cap with %%@
the ${\bf E}\cdot{\bf B} \neq 0$ surface boundary condition (Timokhin %%@
2010) and later with the SCLF boundary condition (Chen \& Beloborodov %%@
2013, Timokhin \& Arons 2013) assuming fixed values of the time-averaged %%@
current density that are either larger than $\rho _{GJ}c$ or of the %%@
opposite sign. These produced evidence for the development of high %%@
potential differences and microsecond bursts of pair creation.

All authors assumed the case ${\bf \Omega}\cdot{\bf B} > 0$ which is not %%@
the subject of the present paper.  Some care will be needed in %%@
translating the results obtained by Chen \& Beloborodov and by Timokhin %%@
\& Arons to the
${\bf \Omega}\cdot{\bf B} < 0$ case. For this, the basic unit of length %%@
in the description of the one-dimensional current and charge density, %%@
derived from the plasma frequency, is almost two orders of magnitude %%@
larger than for electrons.  The one-dimensional potential must form a %%@
continuous function with the three-dimensional potential derived from the %%@
Lense-Thirring effect (see Muslimov \& Tsygan 1992).  In this case, owing %%@
to the larger scale-length, there is no possibility of the particle %%@
backflow noted by Beloborodov (2008) and Chen \& Beloborodov (2013) so %%@
that only the stable zero-temperature solution they refer to is %%@
realistic.  The present paper does not accept the assumption of a %%@
precisely force-free magnetosphere on which the computational work is %%@
based.  Instead, it retains the assumption that the current density is %%@
determined in the polar-cap region and that it flows toward the light %%@
cylinder through a magnetosphere that is not precisely force-free.

Paper IV attempted to describe, qualitatively, the relevance of physical %%@
processes at the surface of the polar cap to the complex phenomena that %%@
are observed in radio pulsars, but a number of numerical estimates made %%@
earlier, in Papers I and II, have been superseded by later work.  Thus %%@
the aim of the present paper is to introduce an elementary mathematical %%@
model of the polar cap which embodies those results obtained in Papers %%@
I-IV that remain relevant. The model itself and the information which has %%@
been derived from it are described in Section 3.  Section 2 summarizes %%@
the physical properties of the polar-cap surface that are essential for %%@
its construction and also considers the non-static aspects of %%@
charge-to-mass ratio fractionation of atmospheric composition which were %%@
neglected in the earlier papers.

It is the scale of the acceleration potential fluctuations which is the %%@
most significant feature displayed by the model.  We relate this with the %%@
observed properties of the RRATs and nulls in Section 4 and with the %%@
phenomena of subpulse drift and null memory in Section 5.  Our %%@
conclusions are summarized in Section 6 along with a brief discussion of %%@
observed phenomena on which, at the present time, our model appears to %%@
have no impact.

\section[]{Essential features of the polar cap}

It is assumed, unlike Papers I and II, that the actual polar-cap magnetic %%@
flux density is of the same order of magnitude as that inferred from the %%@
spin-down rate, which is usually less than the critical field
$m^{2}c^{3}/e\hbar = 4.41\times 10^{13}$ G.  Thus the ion separation %%@
energy (Medin \& Lai 2006) is small enough for the mass of the atmosphere %%@
at a polar-cap temperature $T_{pc}\sim 10^{6}$ K to be significant. 

This atmosphere is very compact and in local thermodynamic equilibrium %%@
(LTE); its scale height is of the order of $10^{-1}$ cm.  The total mass %%@
is poorly known owing to its exponential dependence on the ion separation %%@
energy, but is possibly in an interval equivalent to $10^{-1}$ to %%@
$10^{1}$ radiation lengths. Thus it may contain the whole or some part of %%@
the electromagnetic showers formed by inward accelerated electrons. The %%@
extent of a shower is itself uncertain because the %%@
Landau-Pomeranchuk-Migdal effect is present at the energies, densities %%@
and magnetic fields concerned (see Jones 2010b). At depths immediately %%@
below the atmosphere, the state of matter is uncertain and could be %%@
either liquid or solid (see Paper II).  The nuclear charge is not known, %%@
but we assume the canonical value $Z = 26$.  Electromagnetic showers %%@
reduce this to a mean $Z_{s}$ and mass number $A$ at the top of the %%@
atmosphere with an LTE ion charge $\tilde{Z}$.

\subsection{Atmospheric fractionation}

There is a fractionation of ion charge-to-mass ratio with the largest %%@
values at the top of the atmosphere.  The relatively small number of %%@
protons created in showers are nowhere in equilibrium within the LTE ion %%@
atmosphere and, under the influence of the small electric field $E$ %%@
present, move outward and are either accelerated or, if their flux %%@
exceeds the Goldreich-Julian current density $\rho _{GJ}c$, form an %%@
atmosphere above the ions. Its scale height is $2k_{B}T/m_{p}g$, at local %%@
temperature $T$ and gravitational acceleration $g$. The chemical %%@
potential gradient that causes their motion within the ion sector of the %%@
atmosphere is initially mostly an entropy gradient, but at lower %%@
densities changes to that derived from the electric field.

This fractionation is of central importance to the model described in %%@
this paper, but we have not previously considered the adequacy of our %%@
elementary static treatment of the problem.  Also, the presence of %%@
partial ionization means that the possibility of convective instability %%@
has to be considered (for a simple explanation see, for example, Rast %%@
2001). Bearing in mind the functioning of a laboratory high-vacuum %%@
diffusion pump, it is also necessary to ask if the upward flux of protons %%@
is likely to carry with it sufficient ions to interfere with %%@
fractionation.

In order to examine this, we require two transport relaxation times.
We employ approximate expressions using lowest-order zero-field %%@
perturbation theory, satisfactory here because the proton cyclotron %%@
energy quantum is small compared with the polar-cap thermal energy %%@
$k_{B}T_{pc}$.  
The first, for upward movement of a proton in the ion atmosphere at ion %%@
number density $N_{Z}$ is,
\begin{eqnarray}
\frac{1}{\tau^{tr}_{p}} = \frac{2}{3\pi}\left(\frac{1}{2\pi %%@
m_{p}k_{B}T}\right)^{3/2}N_{Z}\tilde{Z}^{2}e^{4}m_{p}A^{1/2}
\left(A + 1\right)F_{p},
\end{eqnarray}
in which the function $F_{p}$ is,
\begin{eqnarray}
F_{p} & = & \int^{\infty}_{0}\frac{q^{3}dq}{(q^{2} + %%@
\kappa^{2}_{Dp})^{2}}\exp\left(-\frac{q^{2}}{\alpha^{2}_{p}}\right) %%@
\nonumber  \\
& \approx & \frac{1}{2}\ln\frac{\alpha^{2}_{p} + %%@
\kappa^{2}_{Dp}}{\kappa^{2}_{Dp}}.
\end{eqnarray}
Here, $\kappa _{Dp}$ is the Debye wavenumber for the ions,
\begin{eqnarray}
\kappa^{2}_{Dp} = \frac{4\pi e^{2}}{k_{B}T}N_{Z}\tilde{Z}\left(\tilde{Z} %%@
+ 2\right)
\end{eqnarray}
and,
\begin{eqnarray}
\alpha^{2}_{p} = 8m_{p}k_{B}T\frac{A}{(A + 1)^{2}}.
\end{eqnarray}
The equivalent ion relaxation time for movement relative to the proton %%@
atmosphere above the ions, if it exists, is also required.  At proton %%@
number density $N_{p}$ it is,
\begin{eqnarray}
\frac{1}{\tau^{tr}_{Z}} = \frac{2}{3\pi}\left(\frac{1}{2\pi
m_{p}k_{B}T}\right)^{3/2}N_{p}\tilde{Z}^{2}e^{4}m_{p}\frac{A + %%@
1}{A^{2}}F_{Z}
\end{eqnarray}
in which $F_{Z}$ is given by equation (2) with the substitutions,
\begin{eqnarray}
\kappa^{2}_{DZ} = \frac{8\pi N_{p}e^{2}}{k_{B}T},\hspace{1cm}
\alpha^{2}_{Z} = A\alpha^{2}_{p}.
\end{eqnarray}
It is typically several orders of magnitude longer than $\tau^{tr}_{p}$.

The proton drift time from a depth $z = 0$ with ion number density %%@
$N_{Z}$ to the top of the atmosphere is,
\begin{eqnarray}
\tau _{p} = \int^{\infty}_{0}\frac{dz}{\bar{v}(z)} \approx %%@
\frac{l_{Z}}{\bar{v}(0)},
\end{eqnarray}
under the influence of a chemical potential gradient fixed by the ions,
\begin{eqnarray}
m_{p}g - eE = \left(\frac{\tilde{Z} + 1 - A}{\tilde{Z} + 1}\right)m_{p}g,
\end{eqnarray}
equivalent to an upward directed force.  The upward velocity is given by,
\begin{eqnarray}
\frac{1}{\bar{v}(z)} = \frac{\tilde{Z} + 1}{A - \tilde{Z} - 1}
\frac{1}{g\tau^{tr}_{p}(z)},  \nonumber
\end{eqnarray}
and the scale height of the atmosphere is,
\begin{eqnarray}
l_{Z} = \frac{(\tilde{Z} + 1)k_{B}T}{Am_{p}g} \approx 0.15\hspace{2mm} %%@
{\rm cm},
\end{eqnarray}
for $A = 20$, $\tilde{Z} = 6$, $g = 2\times 10^{14}$ cm s$^{-2}$ and $T = %%@
T_{pc} = 10^{6}$ K.  Evaluation for $N_{Z} = 10^{24}$ cm$^{-3}$ gives %%@
$1/\tau^{tr}_{p} = 2.5\times 10^{15}$ s$^{-1}$ and a drift time of $1.0$ %%@
s.  This should be regarded only as a
tentative order of magnitude because equations (1), (3) and (4) are %%@
strictly valid only at number densities much lower than $10^{24}$ %%@
cm$^{-3}$. The position of the change of phase from gas to liquid or %%@
solid is also uncertain.

The proton sector number densities are also sufficient for it to form an %%@
LTE atmosphere, but during phases of proton emission into the %%@
magnetosphere it has the Goldreich-Julian outward flux. Thus as a first %%@
approximation, its kinetic distribution function is isotropic in a frame %%@
of reference moving outward with a velocity $\rho _{GJ}c/N_{p}e$ until %%@
the proton number density $N_{p}$ becomes so small that the LTE condition %%@
breaks down.  If this is not to interfere with fractionation, the force %%@
exerted on an ion by the upward flux of protons must be small compared %%@
with its chemical potential gradient in the static proton LTE atmosphere, %%@
which is,
\begin{eqnarray}
Am_{p}g - \tilde{Z}eE = \left(A - \frac{\tilde{Z}}{2}\right)m_{p}g.
\end{eqnarray}
A proton atmosphere able to maintain a typical Goldreich-Julian flux for %%@
one second would have a density at base of $\sim 10^{22}$ cm$^{-3}$ at %%@
which its ion transport relaxation time would be $1/\tau^{tr}_{Z} = %%@
5\times 10^{10}$ s$^{-1}$ giving a force several orders of magnitude %%@
smaller than equation (10) and of small effect on fractionation.

The existence of convective instability at any density within the %%@
atmosphere is unlikely because the lateral motion implicit in a %%@
convective cell is strongly suppressed by magnetic fields of the order of %%@
$10^{12}$ G.  We refer to Miralles, Urpin \& Van Riper (1997) for a full %%@
treatment of this problem.  But even if convective cells exist, the %%@
presumption must be that the electric field $E$, reflecting the internal %%@
equilibrium of the adiabatically-moving volume, is still present within %%@
it.  Thus the velocity of the protons, averaged over many circulations, %%@
would remain as calculated above and there would be no interference with %%@
fractionation.

\subsection{Acceleration potential}

The polar-cap radius, as in Papers III and IV, denotes the division %%@
between open and closed magnetic flux lines, and is that given by Harding %%@
\& Muslimov (2001),
\begin{eqnarray}
u_{0}(0) = \left(\frac{2\pi R^{3}}{cPf(1)}\right)^{1/2},
\end{eqnarray}
where $P$ is the rotation period.  We assume a neutron-star mass %%@
$1.4M_{\odot}$
and radius $R = 1.2\times 10^{6}$ cm, for which $f(1) = 1.368$. Our %%@
approximation for the electrostatic potential $\Phi$ is  based on the %%@
Lense-Thirring effect described by Muslimov \& Tsygan (1992),
\begin{eqnarray}
\Phi(u,z) = \pi\left(u^{2}_{0}(z) - u^{2}\right) \left(\rho(z) - \rho %%@
_{GJ}(z)\right),
\end{eqnarray}
in cylindrical polar coordinates, at altitude $z$ and radius $u(z)$, for 
the specific case of a charge density $\rho(z)$ that is independent of %%@
${\bf u}$ and is only a slowly varying function of $z$.  (It is almost %%@
identical with the potential that would be present given a %%@
time-independent outward flow of electrons under SCLF boundary conditions %%@
in ${\bf \Omega}\cdot{\bf B} > 0$ pulsars.)
It also assumes approximate forms, valid at altitudes well within the %%@
light-cylinder radius, for the more precise charge densities which were %%@
given by Harding \& Muslimov (2001).  These are independent of ${\bf %%@
u}(z)$ as required by equation (12) and are,
\begin{eqnarray}
\rho _{GJ} = -\frac{B}{cP\eta^{3}}\left(1 - %%@
\frac{\kappa}{\eta^{3}}\right)\cos\psi,
\end{eqnarray}
and
\begin{eqnarray}
\rho = -\frac{B}{cP\eta^{3}}\left(1 - \kappa\right)\cos\psi,
\end{eqnarray}
in which $\psi$ is the angle between ${\bf \Omega}$ and ${\bf B}$, %%@
$\kappa = 0.15$ is the dimensionless Lense-Thirring factor, and $\eta = %%@
(1 + z/R)$. 

Equation (12) is certainly a satisfactory approximation to the true %%@
potential at altitudes $z \gg u_{0}$, but for lower values of $z$, we %%@
must assume that the H-M potential changes continuously to the %%@
one-dimensional potential that exists at $z \ll u_{0}$, which was %%@
originally described by Michel (1974) and then further investigated by %%@
Mestel et al (1985) and by Beloborodov (2008).  We have already noted, in %%@
Section 1, that the length scale associated with the form of the %%@
one-dimensional potential is large, in the ${\bf \Omega}\cdot{\bf B} < 0$ %%@
case. Thus equation (12) is a fair representation of the true potential, %%@
which would remove any possibility of the backflow described by %%@
Beloborodov.

Under the assumption that equation (12) is valid, the condition $\rho(0) %%@
= \rho _{GJ}(0)$ ensures that the SCLF condition ${\bf E}\cdot{\bf B} = %%@
0$ is satisfied at all ${\bf u}$ on the polar-cap surface. The maximum %%@
potential available for acceleration above the polar cap occurs if %%@
$\rho(z)$ is independent of altitude as would be the case if the %%@
particles were ultra-relativistic and there were no charge-separating %%@
interactions.  Expressed in convenient energy units, it is,
\begin{eqnarray}
V_{max}(u,\infty) & = & \frac{2\pi ^{2}R^{3}\kappa eB}{c^{2}f(1)P^{2}} %%@
\nonumber \\  & = &
1.25\times 10^{3}\left(1 - %%@
\frac{u^{2}}{u^{2}_{0}}\right)\frac{B_{12}}{P^{2}} 
\hspace{2mm}{\rm GeV}
\end{eqnarray}
per unit charge on a flux line with radial coordinate ${\bf u}$.

\subsection{Photoelectric transitions}

The blackbody temperatures responsible for photoelectric transitions are %%@
those of the polar cap, $T_{pc}$, and of the whole surface, $T_{s}$.  The %%@
polar cap is not significant at altitudes $z \gg u_{0}(0)$ because the %%@
photon Lorentz transformation to the ion rest frame becomes unfavourable. %%@
There is a little ionization at $z < h \approx 0.05R$ to a mean charge %%@
$Z_{h}$ but the ion Lorentz factors there are small, given the SCLF %%@
boundary conditions, as is the contribution to the total reverse-electron %%@
energy per ion. We assume a fixed value for this component of $\epsilon %%@
_{h} = 20$ GeV. Lorentz transformations of whole-surface photons are much %%@
more favourable at higher altitudes and produce the major part of the %%@
total reverse-electron energy per ion accelerated, $\epsilon = \epsilon %%@
_{h} + \epsilon _{s}$.  Ionization may or may not be complete and we %%@
define the mean final charge as $Z_{\infty}$.  We have shown previously %%@
that the flux of shower-produced protons reaching the top of the %%@
neutron-star atmosphere at any point ${\bf u}$ is given by,
\begin{eqnarray}
J^{p}({\bf u},t) + \tilde{J}^{p}({\bf u},t) = %%@
\int^{t}_{-\infty}dt^{\prime}
f_{p}(t - t^{\prime})K({\bf u},t^{\prime})J^{z}({\bf u},t^{\prime}),
\end{eqnarray}
in terms of the ion flux $J^{z}$ (see equation (20) of Paper IV).  The %%@
first component $J^{p}$ cannot exceed the Goldreich-Julian flux; the %%@
remainder $\tilde{J}^{p}$ accumulates at the top of the atmosphere.  %%@
Paper I assumed the neutron-star atmosphere to be of negligible depth so %%@
that $f_{p}$ was assumed to be the standard diffusion function. But for %%@
the depth considered in Paper IV and here, an expression based on the %%@
drift time given by equation (7) is a better approximation.  If the %%@
proton atmosphere is exhausted, ion emission commences extremely rapidly %%@
in order to satisfy the SCLF condition ${\bf E}\cdot{\bf B} = 0$
at the polar-cap surface.   In general, the screening of any electric %%@
field in the ${\bf \Omega}\cdot{\bf B} < 0$ case for which a positive %%@
charge density is needed occurs preferentially through ion or proton %%@
emission in a single relativistic-particle transit time, whereas the %%@
process of electron-positron pair multiplication requires many transit %%@
times.

Protons accelerated to $\gamma _{p} \sim 10^{3}$ have only a very small %%@
probability of creating electron-positron pairs through interaction with %%@
blackbody photons and, with ions of the same energy per unit charge, %%@
would have only negligible growth rates for the quasi-longitudinal %%@
Langmuir mode.  There are three ways in which particle beams capable of %%@
giving strong coherent radio-frequency emission might be produced.  The %%@
two considered here in the first instance are as follows.

(a) $V_{max}$ is so large that self-sustaining curvature-radiation (CR) %%@
electron-positron pair production occurs. A permanent proton atmosphere %%@
exists over much of the polar cap giving a primary current density of %%@
protons and positrons which, at least for times long compared with %%@
$u_{0}/c$, may be in a steady state. A plasma of low-energy secondary %%@
electrons and positrons forms.

(b) The potential is so reduced from $V_{max}$ by photo-electron backflow 
that ion and proton Lorentz factors are either in a region that allows %%@
rapid growth of the quasi-longitudinal mode or are of magnitude such that 
downward fluctuations to the necessary values can occur. In this case, %%@
the Lorentz factors  are such that the mode wave-vector component %%@
perpendicular to ${\bf B}$ is unlikely to be negligible so that 
coherent radio-frequency emission need not be exactly parallel with local %%@
flux lines.  

The third set of processes are more obscure in the case of ${\bf %%@
\Omega}\cdot{\bf B} < 0$ pulsars and concern inverse Compton scattering %%@
(ICS) of blackbody photons above the polar cap.  Pair production by the %%@
conversion of outward-moving high-energy ICS photons is known to be %%@
significant in ${\bf \Omega}\cdot{\bf B} > 0$ pulsars (see Hibschman \& %%@
Arons 2001, Harding \& Muslimov 2002) even if dipole-field geometry is %%@
assumed.  Also, deviations from such a field, to the extent that they %%@
exist, can greatly enhance pair densities (Harding \& Muslimov 2011).  %%@
But in ${\bf \Omega}\cdot{\bf B} < 0$ pulsars, high-energy ICS photons %%@
are directed inward and the sources of outward-moving positrons to %%@
scatter blackbody photons are more obscure.  Clearly, a fraction of the %%@
photons may have sufficient momentum transverse to ${\bf B}$
to produce pairs at low altitudes.  A further possible source of %%@
low-altitude pairs is the conversion of neutron-capture $\gamma$-rays at %%@
the top of the atmosphere.  Showers produce approximately the same %%@
numbers of neutrons and protons, but reliable calculation of the %%@
probability that a neutron would reach the top of the atmosphere is not %%@
easy.  It is also not obvious how the density of electron-positron pair %%@
production via these processes would respond to fluctuations in the %%@
acceleration potential.  However, although the model described in Section %%@
3 has been considered in the context of either low-energy ion-proton %%@
beams or CR pair production, it is necessary to bear the more obscure %%@
pair-production processes in mind.

\section[]{The polar-cap model and its properties}

The previous Section summarized those specific physical properties of the %%@
polar cap that are important for the model, but we shall also  give a %%@
brief description of its framework.  This is addressed primarily to the %%@
typical radio pulsar with an age of the order of $1$ Myr, or older, and %%@
not to the small-$P$ case in which continuous pair production is %%@
anticipated.

Photoelectric transitions in accelerated ions are a source of electrons %%@
which flow inward to the polar-cap surface and partially screen the %%@
potential given by equation (12). Data presented in Table 1 of Paper IV, %%@
for specific values of $B$ and $T_{s}$, gave $Z_{\infty}$, the final %%@
charge of an accelerated ion, and the energy of its reverse electrons %%@
$\epsilon _{s}$, as functions of the acceleration potential $V({\bf %%@
u},\infty)$ experienced by that ion.  Then with the inclusion of the %%@
small energy $\epsilon _{h}$ arising from photoelectric transitions at %%@
altitudes $z < h$, as described in Section 2.3, the total energy %%@
$\epsilon = \epsilon _{h} + \epsilon _{s}$  gives the number of protons %%@
created in the electron showers.  This is the approximately linear %%@
function $\epsilon W_{p} = KZ_{\infty}$ (see Papers I and II). It is %%@
convenient here to define $K$ as the number of protons formed per unit %%@
ion charge. (In Papers I and II, it was defined as the number of protons %%@
per unit nuclear charge of the ion.) The model described here uses an %%@
elementary form of equation (16) which describes the temporal %%@
distribution of the protons reaching the top of the LTE atmosphere.  But %%@
the distribution of electron back-flow and the values of $Z_{\infty}$ %%@
over the polar-cap also define through solution of Poisson's equation, at %%@
any instant, the actual electrostatic potential which, of course, is not %%@
identical with equation (12) because $\rho$ is not independent of ${\bf %%@
u}$. The model then finds by relaxation procedures 
finite-element approximations to the self-consistent functions
$Z_{\infty}({\bf u},t)$ and $V({\bf u},\infty,t)$ for the polar cap. This %%@
is circular, with radius given by equation (11), which assumption is not %%@
essential but is made for ease of calculation.

\subsection{The model}

The polar cap is divided into $n_{s} = n^{2}$ elements of equal area by %%@
first defining $n$ annuli of outer radius $iu_{0}/n$, where $i = 1....n$ %%@
and then further dividing each of these into $2i - 1$ equal azimuthal %%@
elements. Within each annulus, the elements are displaced, by a random %%@
azimuthal angle, with respect to the axis $\phi = 0$.  It is assumed that %%@
the charge density is independent of ${\bf u}$ within any element.
Equation (16) is much simplified by the model assumption $f_{p}(t) = %%@
\delta(t - t^{\prime} - \tau _{p})$ giving,
\begin{eqnarray}
J^{p}(t) + \tilde{J}^{p}(t) = K(t - \tau _{p})J^{z}(t - \tau _{p}),
\end{eqnarray}
within any element, so that the state of the element alternates between %%@
proton and ion emission.  The total number of protons produced at time %%@
$t$ on unit area of surface at the top of the atmosphere following an %%@
ion-emission interval of length $\tau _{p}$ is,
\begin{eqnarray}
q_{p} = \int^{t - \tau _{p}}_{t - 2\tau _{p}}dt^{\prime}
K(t^{\prime})J^{z}(t^{\prime}).
\end{eqnarray}
The ion current density is $J^{z}(t) = N_{Z}(t)cZ_{\infty}(t)e$ and the %%@
Goldreich-Julian charge density is $\rho _{GJ}(h) = N_{Z}(t)(2Z_{h} - %%@
Z_{\infty}(t))e$.
Therefore, if we define the length of the proton phase by
$\tilde{K}\tau _{p} = q_{p}/\rho _{GJ}c$,
\begin{eqnarray}
\tilde{K}\tau _{p} = \int^{t - \tau _{p}}_{t - 2\tau %%@
_{p}}dt^{\prime}\frac{K(t^{\prime})Z_{\infty}(t^{\prime})}{2Z_{h} - %%@
Z_{\infty}(t^{\prime})}.
\end{eqnarray}
An event is defined as a change of state in any element from proton to %%@
ion emission or vice-versa.  The length of an ion phase is $\tau _{p}$, %%@
which is assumed constant.
The length of the following proton phase is $\tilde{K}\tau _{p}$, and is %%@
a function of the state of the whole polar cap within the integration %%@
time over the preceding ion phase.

Table 1 of Paper IV gave values of $\epsilon _{s}$ and of $Z_{\infty}$ as %%@
functions of a cut-off potential $V_{c}$ for different values of the %%@
surface magnetic flux density $B_{12}$, in units of $10^{12}$ G, and the %%@
whole-surface temperature $T_{s}$.  For reasons which we outline later, %%@
we assume a surface nuclear charge $Z_{s} = 10$.
The cut-off potential is a representation of the effect of photo-electric %%@
transitions on the maximum possible acceleration potential difference %%@
$V_{max}$ given by equation (15).  From Table 1, with some additional %%@
values, we have been able to tabulate the functions $\epsilon _{s}(V)$ %%@
and $Z_{\infty}(V)$ for the interval $0 < V(\infty) < V_{max}$.  They are
not strongly dependent on ${\bf u}$ and we can assume that they are %%@
functions only of the magnitude of $V({\bf u},\infty)$, the potential
at the upper end of the polar-cap acceleration zone which is represented %%@
here by a cut-off at $\eta = 4$. In the model, it is assumed to be the %%@
potential on the central axis of an element.  The excess charge density %%@
within an element during the ion phase is,
\begin{eqnarray} 
\frac{2(Z_{\infty} - Z_{h})}{2Z_{h} - Z_{\infty}}\rho _{GJ}(h),
\end{eqnarray}
and is the source of a potential deviation downward from $V_{max}({\bf %%@
u},\infty)$.  In order to calculate this with some economy, we use the %%@
approximation on which equation (12) is based, specifically that at an %%@
altitude $z >> u_{0}$, a section of the long narrow open magnetosphere %%@
can be approximated locally by a cylinder of constant radius $u_{0}(z)$.  %%@
Then the Green function satisfying the SCLF boundary condition for a line %%@
source at cylindrical polar coordinates $(u, \phi)$ is,
\begin{eqnarray}
G({\bf u},{\bf u}^{\prime}) = \ln\left(\frac{u^{4}_{0} + u^{2}u^{\prime %%@
2} -
2u^{2}_{0}uu^{\prime}\cos(\phi - \phi^{\prime})}{u^{2}_{0}u^{2} + %%@
u^{2}_{0}u^{\prime 2} -
2u^{2}_{0}uu^{\prime}\cos(\phi - \phi^{\prime})}\right).
\end{eqnarray}
The potential on the axis of an element derived from the charge excess %%@
within that element is the most important factor and is obtained by %%@
numerical integration using the Green function. The potential generated %%@
on its axis by any other ion-phase element is found directly from the %%@
Green function by using a line approximation for the excess charge.  The %%@
potential within a proton-phase element is of no interest and is not %%@
calculated.

After each event, the $Z_{\infty}$ in all ion-phase elements, also %%@
referred to as ion-zones, are recalculated by a relaxation procedure so %%@
that there is consistency between the Poisson equation solution for %%@
$V({\bf u},\infty)$ in terms of  all the $Z_{\infty}$ and the tabulated %%@
function $Z_{\infty}(V)$ obtained from Table 1 of Paper IV which is based %%@
purely on the photoelectric transition rates.  This also gives the %%@
function $\epsilon _{s}$ and hence $K$ which, with the values of %%@
$Z_{\infty}$ and their times of duration, then allow calculation of the %%@
integral for $\tilde{K}$ and the duration of the subsequent proton phase.

The initial state of the system has all elements in the proton phase.  %%@
All $n_{s}$ elements, selected in random order, are then assigned %%@
sequential times for transition to the ion phase.  The interval between %%@
any two adjacent times is $x\tau _{p}$, where $x$ is a random number in %%@
the interval $0 < x < x_{max}$.  Our initial choice was $x_{max} = 0.5$.  %%@
Throughout the calculation, a list is maintained, in temporal order, of %%@
the time of the next event in each of the $n_{s}$ elements. This %%@
procedure has the advantage that the calculation of every ion phase is %%@
entirely self-contained so that cumulative errors are not carried %%@
forward.  The model has been run for intervals equivalent to real times %%@
of the order of ten days.

\subsection{Model results}

For a given random initial state, the subsequent states of the model %%@
polar cap are not, of course, random but are chaotic, and variation of %%@
$x_{max}$ appears not to affect their character.  The proton production %%@
parameter $W_{p}$ is a slowly varying function of $B$ but we have chosen %%@
the conservative value $W_{p} = 0.2$ GeV$^{-1}$ and an ion charge $Z_{h} %%@
= 6$ as in Paper IV.  The numbers of elements
have been in the interval $10^{2} \leq n_{s} \leq 4\times 10^{2}$.

The usual procedure has been to let the model run for a time $10^{4}\tau %%@
_{p}$ before sampling its state at intervals of $\tau _{p}$.  The %%@
presence of the instability described in Paper I has been confirmed.  The %%@
system has shown no sign of settling down to other than a chaotic state.  %%@
Its main characteristics are as follows.

(i)		Fluctuations in the central potential $V(0,\infty)$ and in the %%@
number of ion-emission elements can be very large and are dependent, %%@
principally, on the whole-surface temperature $T_{s}$ but also, to a %%@
lesser extent, on the parameter $B_{12}P^{-2}$ which scales $V_{max}$.

(ii)	The elements most likely to be in an ion-emission phase are those %%@
near the periphery $u_{0}$.

(iii)	Peripheral ion-emission elements have a significant %%@
autocorrelation function in the angle $\phi$, unaffected by running time, %%@
indicating the formation of clusters.

At this stage, we should note that the model as precisely defined by %%@
equations (17) - (19) does not produce, at any point on the polar-cap %%@
surface, the mixture of protons and ions which is required for growth of %%@
the quasi-longitudinal Langmuir mode.  But the growth rate is only a %%@
slowly varying function of the proton-ion ratio (see Papers III and IV) %%@
and we do not doubt that, given a physically realistic form of the %%@
diffusion function $f_{p}$ in equation (16), there will be a sufficient %%@
interval of time, of the order of $\tau _{p}$, within which the current %%@
density contains both components.

\begin{table}
\caption{The first two columns give values of the rotation period $P$ and %%@
whole-surface temperature $T_{s}$.  The next five columns histogram the %%@
acceleration potential $\tilde{V} = V(0,\infty)/V_{max}(0,\infty)$  %%@
sampled after $5000$ successive intervals of $\tau _{p}$ in order to show %%@
the scale of the fluctuations that are present in the model for various %%@
$P$ and $T_{s}$. The bin size is $0.2\tilde{V}$ and the number of %%@
elements here is $n_{s} = 100$;  the polar-cap magnetic flux density is %%@
$B_{12} = 3.0$. The final column gives $p_{vn}$, the probability that the %%@
whole polar cap is in the proton phase at any instant.}

\begin{tabular}{@{}rrrrrrrr@{}}
\hline
 $P$ & $T_{s}$ &    &    & $\tilde{V}(0)$ &  &  & $p_{vn}$  \\   

 $s$ & $10^{5}$ K & $0-1 $ & $1-2$ & $2-3$ & $3-4$  
 		& $4-5$  & \\
\hline
	1.0 & 1.0 &	0.000 &	0.004 &	0.008 &	0.021  &  0.967 & 0.000 \\
	    & 2.0 & 0.011 & 0.013 & 0.019 & 0.061 & 0.784 & 0.112  \\
		& 4.0 & 0.018 & 0.064 & 0.144 & 0.302 & 0.464 & 0.008  \\
		
	2.0 & 1.0 & 0.000 & 0.000 & 0.000 & 0.000 & 1.000 & 0.000  \\
	    & 2.0 & 0.000 & 0.000 & 0.001 & 0.146 & 0.854 & 0.000  \\
		& 4.0 & 0.040 & 0.247 & 0.384 & 0.280 & 0.048 & 0.001  \\
		
	3.0 & 1.0 & 0.000 & 0.000 & 0.000 & 0.000 & 1.000 & 0.000  \\
		& 2.0 & 0.000 & 0.000 & 0.004 & 0.388 & 0.608 & 0.000  \\
		& 4.0 & 0.077 & 0.464 & 0.364 & 0.085 & 0.009 & 0.000  \\  
\hline
\end{tabular}
\end{table}

The potential fluctuations given in the Table are, of course, dependent %%@
on $n_{s}$.  The autocorrelation function in $\phi$ indicates that a %%@
smaller of $n_{s}$ should be preferred and, within the limited framework %%@
of our model, would be a better approximation to reality.  But $n_{s} = %%@
100$ is the smallest value for which our approximation for the %%@
interaction between two elements can be reasonably adequate.  Table 1 %%@
therefore underestimates the true scale of fluctuation.
The values of $B_{12}$ and $P$ have been chosen to be representative of %%@
the typical pulsar exhibiting nulls as listed by Wang, Manchester \& %%@
Johnston (2007). The scale of the downward fluctuations in $V$ is, as %%@
might be anticipated, very strongly dependent on $T_{s}$ and hence, %%@
presumably on pulsar age.
At $T_{s} = 10^{5}$ K, significant downward fluctuations would be present %%@
only for very large values of $B_{12}P^{-2}$, the parameter that scales %%@
$V_{max}$.  With increasing rotation period, downward fluctuations also %%@
increase.

The most significant feature of the model is the size of the potential %%@
fluctuations that occur for larger values of $T_{s}$.  But this is %%@
superimposed on a further instability, with medium time-scale, described %%@
in Paper II.  Showers reduce the atomic number from $Z$, possibly $Z = %%@
26$, to $Z_{s}$ at the top of the atmosphere.  Provided we ignore capture %%@
of shower-produced neutrons and subsequent $\beta$-decay of the %%@
neutron-rich nuclei formed, this liberates $Z - Z_{s}$ protons which %%@
diffuse with time-scale $\tau _{p}$. At any instant, there is naturally a %%@
distribution of $Z_{s}$ values, but over a time long compared with the %%@
ablation time (the interval in which one radiation length of matter is %%@
removed from the polar cap at the Goldreich-Julian current density)
\begin{eqnarray}
\tau _{rl} = 2.1\times 10^{5}\left(\frac{-P\sec\psi}
{ZB_{12}\ln(12Z^{1/2}B_{12}^{-1/2})}\right)\hspace{2mm}{\rm s},
\end{eqnarray}
the average quantities must satisfy
\begin{eqnarray}
Z - \langle Z_{s}\rangle  =\langle KZ_{s}\rangle.  \nonumber
\end{eqnarray}
This means that $Z_{s}$ and $\tilde{K}$ defined by equation (19) cannot %%@
be independent variables when considered over medium time-scales of the %%@
order of $\tau _{rl}$.  The reason is that large values of $K$ within %%@
some interval of time imply reduction of $Z$ at the shower maximum to %%@
very small $Z_{s}$ so producing nuclei able to move to the top of the %%@
atmosphere either by diffusion on a time-scale possibly some orders of %%@
magnitude larger than $\tau _{p}$, or by Rayleigh-Taylor instability.  In %%@
particular, values smaller than $Z_{s}\sim 5 $ have a high probability of %%@
being completely ionized in the LTE atmosphere and the consequent absence %%@
of a reverse-electron flux stops shower and proton formation.  The %%@
condition of these surface layers is obviously not well understood, but %%@
in Paper II a case was advanced that it would be one of instability %%@
rather than a steady-state value of $Z_{s}$.  This is the case for our %%@
model assumption of a compromise value $Z_{s} = 10$ and is also the basis %%@
for the observed phenomena, mode-changes and long nulls, with time-scales %%@
of the order of $\tau _{rl}$.

\section[]{RRATs nulls and potential fluctuations}

Our proposal is that the potential fluctuations found in the model %%@
provide a basis for understanding why the extent of null lengths and %%@
fractions seen in both pulsars and the RRATs occur quite naturally.

\subsection{The RRATs}

We refer to Burke-Spolaor \& Bailes (2010), Keane et al (2011) and Keane %%@
\& McLaughlin (2011) for reviews of the observational data on these %%@
sources.  Values of $f_{on}$, the fraction of periods in which a pulse is %%@
detected, are broadly in the interval $10^{-4} < f_{on} < 10^{-1}$ within %%@
which their distribution is  approximately uniform in $\log f_{on}$.  The %%@
bursts of emission are short, usually one but occasionally several %%@
periods, and are consistent with being of the same order of magnitude as %%@
$\tau _{p}$. The RRATs listed in Table 3 of Keane et al are all quite %%@
distant with the exception of J1840-1419 which also has a period $P = %%@
6.6$ s and a value of the acceleration potential parameter $B_{12}P^{-2} %%@
= 0.15$, that is, below the cut-off value $0.22$ which can be found from %%@
the ATNF catalogue (Manchester et al 2005).

Thus Weltevrede et al (2006a) proposed that the RRAT pulses are simply %%@
analogues of the giant pulses seen in the nearby pulsar B0656+14 and that %%@
their emission, apart from the giant pulses, is unobservable owing to %%@
distance.  However, there is a difference whose significance becomes %%@
obvious in view of the model described here.  With the exception of %%@
J1554-5209, the $14$ RRAT listed with surface magnetic fields by Keane et %%@
al are well below the threshold for CR secondary pair production as %%@
defined by Harding \& Muslimov (2002).  From Fig.1 of their paper, this %%@
condition can be represented approximately in terms of the parameter
$X = B_{12}P^{-1.6}$: the threshold is given by the critical value $X_{c} %%@
= 6.5$.   PSR 0656+14 lies well above this threshold, which assumes %%@
dipole-field flux-line curvature, as does the exceptional RRAT %%@
J1554-5209, and it is reasonable to suppose that these pulsars are %%@
capable of supporting CR secondary pair creation over at least a fraction %%@
of their polar caps. Therefore, we suggest that the giant pulses of %%@
B0656+14 are produced by process (a) through a rare upward fluctuation in %%@
$V$ toward $V_{max}$ and its more general emission by process (b).   Some %%@
RRATs may, of course, have sufficient non-dipole flux-line curvature to %%@
enable process (a), but discounting this possibility, the RRATs listed by %%@
Keane et al are so far below the Harding-Muslimov  threshold that we %%@
suggest their isolated pulses are a consequence of rare but large %%@
downward fluctuations in $V$.

It might be argued that, owing to the boundary condition satisfied by %%@
equation (12), Lorentz factors $\gamma _{A,Z}$ small enough for growth of %%@
the quasi-longitudinal Langmuir mode always exist, in principle, at ${\bf %%@
u}$ close to $u_{0}$ so that there should be some coherent emission at %%@
all times.  But in this case it is not obvious that the lateral %%@
(perpendicular to ${\bf B}$ and to ${\bf u}$) depth of the particle beam %%@
would be suitably large in relation to the wavelength of the mode to %%@
allow the necessary growth rate. There is an obvious need for some %%@
analogue of the relativistic Penrose condition (see Buschauer \& Benford %%@
1977) extended to allow  for a beam with lateral velocity gradients and %%@
limited lateral depths.   The question of the conditions necessary for %%@
observable emission will be further considered in Section 5.1 in relation %%@
to the structure and drift of subpulses.

The ages and surface magnetic fields of RRATs given by Keane et al can be %%@
compared with those of the nulling pulsars listed in Tables 1 and 2 of %%@
Wang, Manchester \& Johnston (2007).  It is immediately seen that RRAT %%@
fields are typically an order of magnitude larger than those given in the %%@
ATNF catalogue (Manchester et al 2005) for the Wang et al list.  But the %%@
distributions of the acceleration parameter $B_{12}P^{-2}$ are very %%@
similar because the RRAT have long periods.   The age distributions are %%@
also similar though both are very wide.

The relation between age and proper-frame surface temperature $T_{s}$ is, %%@
unfortunately, obscure in the region assumed in Table 1, which is well %%@
within the photon-cooling epoch.  Potekhin \& Yakovlev (2001; Fig.7) have %%@
shown that increasing surface fields accelerates cooling, but an even %%@
more important factor would be the presence of low-$Z$ elements in the %%@
outer crust, possibly from fall-back, which have the same effect %%@
(Potekhin et al 2003).

\subsection{Null lengths and surface temperature}

With reference to Table 1, it is possible to see qualitatively how nulls %%@
change with age during the cooling from high-$T_{s}$ to low-$T_{s}$.  We %%@
consider pulsars which lie below the pair-creation threshold.  Initially, %%@
values of $\tilde{V}$ are usually small enough to give the mode growth %%@
rates necessary for emission by process (b) (see Section 2.3).  Upward %%@
fluctuations of $\tilde{V}$ increase the ion and proton Lorentz factors %%@
and, owing to the exponential dependence of  amplitude growth on them,  %%@
can produce short nulls.  Upward fluctuations become more frequent as %%@
cooling proceeds, giving higher null fractions $1 - f_{on}$. Eventually, %%@
cooling reaches the stage at which the mean $\tilde{V}$ approaches unity %%@
and downward fluctuations are required to produce the growth rates %%@
necessary for observable emission.  Null fractions are then near unity %%@
and, as the limit is approached, there is no observable emission.  We %%@
emphasize that these fluctuations are superimposed on medium time-scale %%@
fluctuations (see Section 3.2) in the surface atomic number $Z_{s}$ which %%@
are unlikely to be uniform over the whole polar-cap surface.  Values %%@
smaller than $Z_{s} \sim 5$ produce a negligible reverse-electron energy %%@
flux and hence, if present over a substantial area of the polar cap, much %%@
reduced deviations of $\tilde{V}$ from unity.  It is proposed here that %%@
these are responsible for null and burst lengths more nearly of the order %%@
of $\tau _{rl}$ rather than $\tau _{p}$.

Isolated Neutron Stars (INS) are a small group of radio-quiet thermal %%@
X-ray emitting sources positioned close to the RRAT in the $P-\dot{P}$ %%@
plane (see Keane et al 2011) but at slightly greater periods. The six %%@
sources for which timing solutions exist have been listed by Zhu et al %%@
(2011). Given their kinematic ages, the observer-frame temperatures are %%@
large ($\sim 10^{6}$ K) compared with those considered in Table 1 but, %%@
with the exception of the anomalous J0420-5022, their $B_{12}P^{-2}$ are %%@
typically close to the general cut-off value of 0.22 found from the ATNF %%@
catalogue (Manchester et al 2005). If they form part of the ${\bf %%@
\Omega}\cdot{\bf B} > 0$ population, they should support pair formation %%@
by the ICS mechanism; for ${\bf \Omega}\cdot{\bf B} < 0$, extrapolation %%@
from Table 1 shows that small $\tilde{V}$ would be expected satisfying %%@
the condition necessary for rapid growth of the ion-proton beam %%@
instability.  However, the fact that no radio emission has been observed %%@
from this small number of sources is, perhaps, not surprising in view of %%@
their anticipated small beaming fractions compared with the %%@
$4\pi$-observability of the whole-surface X-rays.

\section[]{Subpulse drift and null memory}

Subpulse drift, and the null memory which, in some pulsars, is observed %%@
with it, is obviously an important diagnostic of polar-cap physics.  %%@
Unfortunately, our model is incomplete in that it does not have the %%@
capacity to spontaneously exhibit this phenomenon following the %%@
randomly-constructed initial state described in Section 3.  Its %%@
finite-element structure and the very elementary approximation %%@
represented by equation (17) are the reasons for this.  However, the %%@
phenomenological model developed by Deshpande \& Rankin (1999) from the %%@
classic Ruderman \& Sutherland (1975) model has proved so useful that any %%@
physical polar-cap model must be able to support it.

\subsection{Nulls and null memory}

It was noted in Section 3 that the model system settles down to a chaotic %%@
state, but with ion-emission elements most likely to be near the %%@
periphery $u_{0}$ and an autocorrelation function in the azimuthal angle %%@
$\phi$ indicating the formation of clusters.  This is unsurprising, %%@
because ion-emission elements at smaller $u$ have larger reverse-electron %%@
energy fluxes and so lead to longer phases of proton emission.    
Thus our model generates quite naturally the conal structure that is %%@
required for the Deshpande-Rankin carousel model and we can therefore %%@
proceed to examine the extent to which it is capable of supporting its %%@
organized subpulse motion.

For convenience, we shall consider a circular path of radius $u$ on the %%@
polar cap and associate a moving ion zone with the formation of a %%@
subpulse.  Then it is possible to envisage organized circular motion in %%@
which an ion-emission zone of angular width $\delta\phi _{i}(u)$ is %%@
followed by a proton-emission zone of $\tilde{K}\delta\phi _{i}(u)$.  %%@
Thus at any instant,
\begin{eqnarray}
\sum _{i}\delta\phi _{i}(u)\left(1 + \tilde{K}_{i}(u)\right) = 2\pi,
\end{eqnarray}
with $i = 1 .... n$ in which $n$ is here the number of elements on the %%@
carousel and each $\tilde{K}$ is determined according to equation (19) by %%@
the preceding ion-zone.  Comparison with measured values of the %%@
longitudinal subpulse separation $P_{2}$ indicates that the order of %%@
magnitude of model values should not be large, $\tilde{K}_{i}\sim 3$, but %%@
this order of magnitude is associated with values of the acceleration %%@
potential, on the ion-zone flux lines, that are unlikely to enable %%@
significant electron-positron pair creation.  This is then consistent %%@
with mechanism (b), based on protons and ions, for the production of %%@
coherent radio emission.

It is easy to see that a steady uniform state of equation (23) with 
$\delta\phi = \tau _{p}\dot{\phi}$ and constant $\dot{\phi}$ that is %%@
independent of $u$
does not, in general, exist.  However, we have to bear in mind the %%@
limitations of our model based on equation (16), which is local in ${\bf %%@
u}$, and on its very elementary approximation given by equation (17).  %%@
Thus in a model with a more realistic diffusion function $f_{p}$ and %%@
without the finite element structure, we would expect that at constant %%@
$\phi$, $\tilde{K}$ would increase as a function of $u$, starting from a %%@
negligible value at $u_{0}$ and reaching a maximum before declining as a %%@
proton-emission zone is entered.  The presence of this extremum is a %%@
possible basis for the existence of short-term quasi-steady-state %%@
solutions of equation (23) within a finite interval of $u$ at $u_{c}$ %%@
which also coincides with conditions suitable for the growth of the %%@
instability giving observable emission.  It may be that the tendency to %%@
form a cluster, present in our model, indicates that a more realistic %%@
model would counteract the dispersion inherent in equation (23) and %%@
maintain subpulse shape, but it is not possible to assert that it would %%@
be so. 

However, it is possible to envisage this motion on a polar cap of any %%@
plausible geometrical shape, elliptical or approximately semi-circular.
The motion is, of course, independent of ${\bf E}\times{\bf B}$ drift and %%@
could be in either direction, as is seen in the survey of Weltevrede, %%@
Edwards \& Stappers (2006b), or may even be bi-directional at some %%@
instant on different areas of the polar cap. There are many ways in which %%@
potential fluctuations could disrupt this organized motion. We have seen %%@
in the previous Section that a whole-surface temperature $T_{s}$ large %%@
enough to give low values of $\tilde{V}$ favours the occurrence of nulls %%@
through upward potential fluctuations.  The most simple fluctuation is %%@
one that increases the potential over much of the polar cap and in %%@
particular at $u=u_{c}$. The functions $\epsilon _{s}(V)$, and hence %%@
$\tilde{K}$, increase quite rapidly with $V$.  There are two %%@
consequences.  Firstly, the exponent in the expression for the mode %%@
growth rate (Paper IV, equation 17) is dependent on $\gamma %%@
_{A,Z}^{-3/2}$ so that the conditions necessary for observable coherent %%@
radio emission are not reached and a null occurs.  Secondly, the %%@
(unobserved) motion of the ion zones is substantially changed. 

It is possible to see the nature of these changes by considering how two %%@
variables, the proton density $q(\phi,t)$ at the top of the atmosphere %%@
and the ion current density $J^{z}$ satisfying equations (17) - (19) 
on a section of the path at $u = u_{c}$, evolve over a sequence of times.
In doing this, we assume that the ion-zone is able to move continuously %%@
and abandon the fixed finite-element calculation of the model described %%@
in Section 3.1  The ion-zone motion during a null shows how a form of %%@
null memory occurs.

It is convenient, for this problem, to represent the density $q$ in units %%@
of $\rho _{GJ}(0)c\tau _{p}$ and $J^{z}$ in units of $\rho _{GJ}(0)c$.
The ion-zones at time $t$ have $J^{z}\sim 1$, width $\delta\phi^{(1)}$ %%@
and move with velocity $\dot{\phi}^{(1)} = \delta\phi^{(1)}/\tau _{p}$.
Each generates a proton density $q(\phi,t)$ at the top of the atmosphere %%@
which is depleted by the  proton-zone current density $J^{p} = 1$.  Thus %%@
at an instant $t$, and as a function of $\phi$, the proton density %%@
following an ion zone rises linearly to a maximum of $\tilde{K}^{(1)} - %%@
1$ in a distance $\delta\phi^{(1)}$ and falls linearly to zero in a %%@
further distance $(\tilde{K}^{(1)} - 1)\delta\phi^{(1)}$.  The velocity %%@
of an ion zone is therefore determined by the gradient $\partial %%@
q/\partial\phi$ of the preceding proton zone.
This is the quasi-steady-state motion at time $t$, at which instant, an %%@
upward potential fluctuation increases proton production to %%@
$\tilde{K}^{(2)}$.
The increase in ion Lorentz factor reduces the quasi-longitudinal mode %%@
growth rate to a sub-critical value and so initiates the null.  A new %%@
maximum proton density $q = \tilde{K}^{(2)} - 1$ is reached at a later %%@
time $t + 2\tau _{p}$ but the system velocity remains $\dot{\phi}^{(1)}$
until $t + \tilde{K}^{(1)}\tau _{p}$.  The width of an ion zone is then %%@
reduced.  In detail, and with reference to equations (17) - (19), this %%@
happens because the velocity of its leading edge is reduced whilst that %%@
of the rear edge remains at $\dot{\phi}^{(1)}$ until
$t + (\tilde{K}^{(1)} + 1)\tau _{p}$ at which time the ion-zone width and %%@
velocity are both reduced by a factor of
$(\tilde{K}^{(2)} - \tilde{K}^{(1)} + 1)$.
The ion-zone velocity then tends to its new steady-state value %%@
$\dot{\phi}^{(2)} =
\dot{\phi}^{(1)}(1 + \tilde{K}^{(1)})/(1 + \tilde{K}^{(2)})$.  Hence %%@
subpulse drift does not stop in the model, but is much reduced after a %%@
null time interval of
$\tilde{K}^{(1)}\tau _{p}$.  Thus for short nulls, there is a subpulse %%@
memory, but it does not correspond precisely with the conservation of %%@
subpulse longitude over a short null that has been observed, for example %%@
in PSR B0809+74, by van Leeuwen et al (2002).

We are not unduly disturbed by this, because there are many possible %%@
spatial distributions of potential fluctuation over the polar cap other %%@
than the uniform case considered above.  It is also not clear that the %%@
precise form of memory seen in B0809+74 is universally observed.
The upward potential fluctuation that caused this could reverse at any %%@
time during or after the sequence described in the previous paragraph.  %%@
In relation to null memory, the question then is, when does observable %%@
coherent radio emission recommence? The problem arises owing to the very %%@
elementary nature of our model defined by equations (17) - (19) and is %%@
that $\delta\phi^{(2)}$ in the model can be very small.  The acceleration %%@
potential within a very small interval of $\phi$ may be too large to %%@
permit adequate growth of the mode and the conditions discussed briefly %%@
in Section 4.1 may not be satisfied.

\subsection{Subpulse drift}

Published observations on mode-changes and subpulse drift demonstrate the %%@
extremely heterogeneous nature of these phenomena.  Some features are %%@
quite distinct in a small number of pulsars, but much less distinct or %%@
not present at all in others.  For this reason, we have used the results %%@
of the extensive survey made by Weltevrede et al (2006b) based on %%@
observations of 187 pulsars selected only by signal-to-noise ratio, and %%@
list some of their conclusions below.

(i)   Roughly equal numbers of pulsars have subpulse drift to smaller or %%@
greater longitudes, and direction reversal is observed.

(ii)  Bi-directional drifting is observed in a small number of pulsars. %%@
J0815+09 and B1839-04 have mirrored drift bands.

(iii) The drift rate can be mode-dependent, having one of a number of %%@
discrete values of the band separation $P_{3}$, and nulls may also be %%@
confined to a particular drift-mode.

(iv)  Drift bands can be curved or non-linear with pulse %%@
longitude-dependent spacing of subpulses: they are often indistinct and %%@
can be found only by two-dimensional spectral analysis.

(v)   The band separation $P_{3}$ is uncorrelated with age %%@
($P/2\dot{P}$), or with $P$ and $B$.

Conclusions (i) and (ii) are quite naturally consistent with the model.  %%@
We emphasize again that ${\bf E}\times{\bf B}$ drift is not involved and %%@
there is no reason why, given the chaotic state of the polar cap, there %%@
should not be reversals or even bi-directionality. The nature of the %%@
chaotic state also means that it is unrealistic to suppose that these %%@
phenomena can be predictable.  Conclusion (iii) is also a natural %%@
consequence of medium time-scale instability in the value of the surface %%@
atomic number $Z_{s}$ and hence in the potential as mentioned in Section %%@
4.2.   The existence of a polar-cap area emitting nuclei with $Z_{s}$ too %%@
small to produce a significant reverse-electron flux is obviously %%@
associated with an upward displacement of $\tilde{V}$ for times very %%@
broadly of the order of $\tau _{rl}$.  This is quite consistent with the %%@
observed time intervals of $10^{3-4}$ s for mode-changes.
Non-linear drift bands (iv) are simply a consequence of non-uniformity of %%@
$\tilde{K}_{i}$ values in equation (23).  Band separation $P_{3}$ was %%@
discussed in Paper II.  It is given by $P_{3} = (\tilde{K} + 1)\tau %%@
_{p}$.  In this expression, $\tau _{p}$ is dependent only on the %%@
properties of the LTE atmosphere at the polar-cap surface, but %%@
$\tilde{K}$ may have some dependence on rotation period and magnetic flux %%@
density; also on whole-surface temperature $T_{s}$ and hence pulsar age.
It was noted in Paper II that the distribution of $P_{3}$ is quite %%@
compact, most of the values listed by Weltevrede et al (2006b) being %%@
within the interval $1 < P_{3} < 10$ s.  The fact that the carousel path %%@
is restricted to being near the polar-cap periphery $u_{0}$ may well act %%@
as a constraint on the values of $\tilde{K}$ and so $P_{3}$ that occur.

Finally, Kloumann \& Rankin (2010) have observed randomly distributed %%@
pseudo-nulls in B1944+17 of length less than $7P$ with weak emission, but %%@
well above noise. They interpreted them as a state of no subpulse on that %%@
band of polar-cap flux lines from which photons can enter the line of %%@
sight. These occur naturally in the model.  The final column of Table 1 %%@
gives $p_{vn}$, the probability that the whole polar cap is in the proton %%@
phase at any instant.  There would then be no possible way of producing %%@
particle beams satisfying conditions (a) or (b) of Section 2.3 and %%@
capable of generating coherent radio emission. This contributes to the %%@
total null fraction that is observed. The probability $p_{sn} > p_{vn}$ %%@
that there will be no ion-zones merely in the part of the polar cap which %%@
is observable along the line of sight is therefore a natural parameter of %%@
the model.

\section[]{Conclusions}

Papers I - IV, followed by the present paper, have attempted to predict %%@
some of the consequences of there being two populations of isolated %%@
pulsars having opposite spin directions. In many important respects, the %%@
physical state of  the polar-cap has been found to be dependent on the %%@
sign of the Goldreich-Julian charge density.  In the ${\bf %%@
\Omega}\cdot{\bf B} < 0$ case considered in these papers, SCLF boundary %%@
conditions have been assumed rather than the ${\bf E}\cdot{\bf B}\neq 0$ 
condition of the original Ruderman \& Sutherland (1975) model. Then the %%@
state of the polar cap is unstable on time-scales of the order of $\tau %%@
_{p}$ and of the longer polar-cap ablation time $\tau _{rl}$.  These are %%@
both many orders of magnitude longer than the characteristic polar-cap %%@
time-scale of $u_{0}/c$ and there can be little doubt that the system %%@
should be able to allow the movements of charge necessary to maintain the %%@
SCLF boundary conditions for them.  We therefore need to look at %%@
published observational data to see if there is any real evidence for the %%@
existence of two populations.

The Weltevrede et al (2006b) survey is an obvious starting point.  It %%@
lists 187 pulsars selected only by signal-to-noise ratio and examines %%@
both subpulse modulation (the wide variations of intensity at a fixed %%@
longitude in a sequence of observed pulses) and subpulse drift. Subpulse %%@
modulation is an almost universal characteristic and we assume it to be a %%@
direct consequence of the plasma turbulence that itself is now widely %%@
believed to be the source of the radio emisssion (see Asseo \& Porzio %%@
2006). Subpulse drift with measurable values of both longitudinal %%@
separation of successive subpulses $P_{2}$ and band separation $P_{3}$
was detected in 72 pulsars and these can be compared with a set of 113 %%@
which do not show detectable subpulse drift.  The distributions of both %%@
sets as a function of age are wide and are broadly similar except that %%@
only 7 of the 72 have an age less than 1 Myr as opposed to 31 of the set %%@
of 113.  The distributions as functions of the parameter $X$ whose %%@
critical value $X_{c} = 6.5$ defines the CR pair creation threshold are %%@
also wide but the set of 113 has an excess at $X > X_{c}$.  There are 35 %%@
pulsars with $X > X_{c}$ as opposed to only 7 of the 72.  This is %%@
consistent with, but does not prove, the existence of two populations %%@
which separate as $X$ moves with age to values below $X_{c}$. We propose %%@
that the ${\bf \Omega}\cdot{\bf B} < 0$ set show the phenomena of %%@
mode-changes, nulls and subpulse drift as they age through growth of the %%@
ion-proton beam quasi-longitudinal mode (mechanism (b) of Section 2.3 but %%@
see also the comments there on more obscure processes of pair creation %%@
for pulsars of this spin direction).  Pair creation by ICS photons in the %%@
${\bf \Omega}\cdot{\bf B} > 0$ case has been very fully investigated by 
Hibschman \& Arons (2001) and by Harding \& Muslimov (2002, 2011) and our %%@
assumption is that this is the source of coherent emission in the %%@
population that has no nulls or subpulse drift.

Having divided the Weltevrede et al list into two populations, we should %%@
compare these with the extensive survey of nulls made by Wang et al %%@
(2007).  No pulsar in Table 1 of Wang et al appears in the earlier %%@
Weltevrede et al list, presumably because they were not known or did not %%@
satisfy the selection criteria. Of the 46 pulsars with measured null %%@
fractions given in Table 2 of Wang et al, 18 also do not appear in %%@
Weltevrede et al, but 20 are listed with a measured value of $P_{3}$ and %%@
8 have no detected $P_{3}$.  A smaller but more recent list of nulling %%@
pulsars in the paper of Gajjar, Joshi \& Kramer (2012) includes a further %%@
8 that were not considered by Wang et al, of which 2 have measured values %%@
of $P_{3}$ but 6 do not appear in the paper of of Weltevrede et al.  Of %%@
the 8 pulsars that have nulls but no detected $P_{3}$, 4 have null %%@
fractions $1 - f_{on} < 0.01$ and B0656+14 is the special case described %%@
in Section 4.1. There are only 3 pulsars  having substantial null %%@
fractions but no detected $P_{3}$. These are the otherwise unremarkable  %%@
B1112+50, B2315+21 and B2327-20.  But it is not necessarily correct to %%@
regard them as anomalous because, as noted by Weltevrede et al, drift %%@
bands are often indistinct and detectable only by two-dimensional %%@
spectral analysis.  We suggest that the results of this comparison are by %%@
no means inconsistent with our division of the Weltevrede et al pulsars %%@
into two populations. 

A different, but perhaps more anomalous set are the three pulsars known %%@
to exhibit nulls of very long duration, of the order of $10$ d.  The %%@
first of these to be found (B1931+24; Kramer et al 2006) enabled %%@
spin-down rate measurements to be made separately in both on and off %%@
states of emission. This was followed by similar measurements for %%@
J1832+0029 (Lorimer et al 2012) and J1841-0500 (Camilo et al 2012).  In %%@
each case, the off-state spin-down rate was about half that of the on %%@
state.  But these pulsars have quite large values $X = 3.6$, $2.5$ and %%@
$6.6$ respectively, close to the critical value $X_{c} = 6.5$, and may %%@
quite possibly support self-sustaining CR pair creation during the on but %%@
not the off state.  The consequent difference in the flux and nature of %%@
particles passing through the light cylinder appears to be the only %%@
physically plausible mechanism for a spin-down torque change of this %%@
magnitude.  But the $10^{6}$ s time-scale for both on and off states of %%@
emission appears  a little too long compared with times of the order of %%@
$10\tau _{rl}$ found from equation (22).  The question of %%@
quasi-periodicity with such time-scales should also be addressed, as in %%@
the case of the RRATs (Palliyaguru et al 2011).  We have stressed that %%@
our model is not random, but deterministic.  Quasi-periodicites are
therefore not impossible in principle, but remain quite difficult to %%@
explain. 

Changes in the polar-cap acceleration potential have little direct effect %%@
on the spin-down torque. Changes in the proton-ion composition occurring %%@
in the process (b) nulls described in section 4.2 would lead to a change %%@
in the mean charge to mass ratio of particles crossing the light %%@
cylinder, but it must be doubted whether mechanism (b) could explain the %%@
large observed difference in spin-down torque. However, it is not yet %%@
clear that this is a universal feature of nulls.  The only other pulsar %%@
for which measurements have been made (PSR B0823+26: Young et al 2012) %%@
has a fractional upper limit of $0.06$ for the change in spin-down %%@
torque.  It has a value $X = 2.6$, rather below the putative critical %%@
value $X_{c} = 6.5$ denoting the pair creation threshold, so that the %%@
relatively small change in torque could be consistent either with %%@
relatively weak pair production or with process (b).

In the absence of electron and positron densities large enough to adjust %%@
and so cancel an electric field, as in mechanism (a), we have to remember %%@
that acceleration or deceleration remains at higher altitudes beyond %%@
$\eta \sim 10$ as a consequence of natural flux-line curvature.  But at %%@
this stage, growth of the quasi-longitudinal mode has already occurred so %%@
that further acceleration would have negligible effect on the coherent %%@
radio emission.  This is a further distinction between the two %%@
populations.  In the ${\bf \Omega}\cdot{\bf B} < 0$ case, the polar cap %%@
of open magnetic flux lines may well have an approximately semi-circular %%@
shape.  In this case, the observed drift of subpulses would be along a %%@
diameter instead of an arc of a circle.

Although all the above problems remain, the model ${\bf \Omega}\cdot{\bf %%@
B} < 0$ polar cap described here has some positive features. It does not %%@
require that neutron-star magnetic fields lie in a particular interval.  %%@
This is important because radio pulsar inferred polar fields can vary by %%@
up to six orders of magnitude.  The physical processes in electromagnetic %%@
shower development or in the photoelectric transitions exist in the %%@
zero-field limit and do not change in any qualitative way with increasing %%@
field.

The model is deterministic, but chaotic, but is incomplete in that its %%@
use of finite elements for the reason stated at the end of Section 3.1 %%@
and the very elementary nature of the approximation made in equation (17) %%@
appear to preclude the spontaneous appearance of subpulse drift from the %%@
random initial state which is used.  It is possible to do no more than %%@
assert that the model can support subpulse drift in a quasi-stable way.

It is unfortunate that quantitative model predictions depend so much on %%@
surface atomic number and particularly on whole-surface temperature %%@
$T_{s}$, parameters which are not well-known.  Cooling calculations (see %%@
the review of Yakovlev \& Pethick 2004) show that $T_{s}$ falls steeply %%@
at ages greater than 1 Myr.
Bearing in mind that, for a neutron-star with the mass and radius assumed %%@
here, the observer-frame temperature is $T^{\infty}_{s}\approx 0.8T_{s}$,
we can see that the temperatures assumed in Table 1 fall well below the %%@
values currently observable. Although cooling in this interval is %%@
photon-dominated, the whole-surface temperature must be regarded as very %%@
uncertain.

But having made these reservations, the model does represent a %%@
physically-realistic framework for understanding the reasons why RRATs %%@
and the varied phenomena of mode-changes, nulls and subpulse drift appear 
during neutron-star aging.  It may be that further observations at %%@
frequencies below 100 MHz will provide evidence for the existence of a %%@
population emitting by process (b) and having spectra biassed towards  %%@
lower frequencies than those for the process (a) population.

\bsp

\label{lastpage}

\end{document}